\documentclass[conference,10pt]{IEEEtran}

\usepackage{epsfig,rotating,setspace,latexsym,amsmath,epsf,amssymb,bm,theorem}
\usepackage{cite}

\newtheorem{Theorem}{Theorem}

\newtheorem{remark}{Remark}
\newenvironment{Proof}[1]{\medskip\par\noindent
{\bf Proof:\,}\,#1}{{\mbox{\,$\blacksquare$}\par}}

\def \n2{{N_0 \over 2}}

\def \h5{\hspace{0.5in}}

\begin{document}
\IEEEoverridecommandlockouts
\pagestyle{empty}

\title{Timely Status Updating Through Intermittent Sensing and Transmission \vspace{-0.1in}}

\author{Omur Ozel \\
\normalsize Department of Electrical and Computer Engineering \\
\normalsize George Washington University, Washington, DC 20052 USA \\
\normalsize {\it ozel@gwu.edu}\vspace{-0.2in}}

\maketitle 

\begin{abstract}
We consider a novel \textit{intermittent status updating} model where an energy harvesting node with an intermittent energy source performs status updating to a receiver through \textit{non-preemptive} sensing and transmission operations. Each operation costs a single energy recharge of the node and the node cannot harvest energy while in operation. The sensing time for each update is independent with a general distribution. The transmission queue has a single server receiving packets generated after sensing operation, general service time distribution and a single data buffer to save the latest arriving update packet. Once energy is harvested, the node has to decide whether to activate sensing to generate a new update or transmission to send the existing update (if any) to the receiver. We prove that average peak age of information (AoI) at the receiver is minimized by a threshold-based stopping rule that accepts only \textit{young} packets to the transmission server. We then use this result to address average AoI optimization over the considered stopping rules through novel hybrid waiting and thresholding schemes. Our numerical results show the improvements in average AoI maintained by hybrid schemes. 
\end{abstract}


\section{Introduction}

This paper considers a novel \textit{intermittent status updating} model motivated by intermittently powered energy harvesting systems where operations are performed one by one as new energy is replenished and execution order must take this intermittency into account (see e.g. \cite{hester2017timely,lucia2017intermittent}) with minimal to no energy storage. Once power is restored after a power outage, the device has to decide whether to finish current execution for forward progress or to start again with a fresh status update. Our focus will be exclusively on sensing and transmission operations while their representations as servers and queues apply more generally. We use AoI metric for timely updating in an energy harvesting node with no battery which necessitates allocating energy to sensing or transmission at the time of arrival. No energy can be harvested during operation.

AoI metric has received extensive research interest as a measure of staleness of available information at monitoring receivers of a system. Since the pioneering works in \cite{kaul2012status, kaul2012real} for various queuing models, the AoI metric has been used for timely information updating models and applications \cite{bedewy2017age, talak2017minimizing, yates2018age, maatouk2018age,Alabbasi2018JointIF,Gong2019ReducingAF,xu2019peak,pimrc19}. Of particular interest and relevance to our current work are the papers performing AoI analysis and optimization in energy harvesting systems \cite{bacinoglu2015age, yates2015lazy, wu2017optimal_ieee, arafa2017age,farazi2018average,farazi2018age, feng2018optimal, bacinoglu2018achieving,2018information,leng2019age,ceran2019average}. In another related line of research, \cite{costa2016age} investigates the role of packet management to improve the average AoI at the monitoring node. \cite{inoue2018general} provides a general treatment of stationary probability analysis of AoI in various preemptive and non-preemptive queuing disciplines. Reference \cite{kam2018age} considers introducing packet deadlines to discard the packets in a single server system for timeliness. \cite{kavitha2018controlling} addresses the problem of packet drop control for information freshness. In \cite{sun2017update,infocom_arxiv,infocom_w,bedewy2019age} \textit{waiting} is used as a mechanism to regulate the traffic for improving average AoI with increased peak AoI cost. Benefits of waiting are also considered for sampling a Wiener process for remote estimation in \cite{yinsun}.

\begin{figure}[!t]
\centering{
\hspace{-0.0cm} 
\includegraphics[totalheight=0.15\textheight]{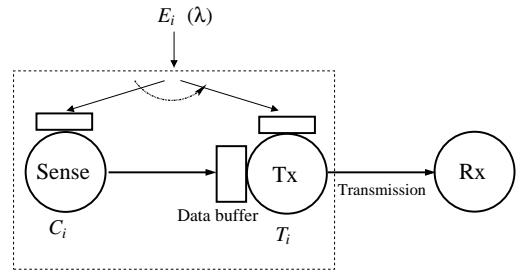}}\vspace{-0.1in}
\caption{\sl System model for an EH node performing sensing and transmission operations by energy harvested from the environment.}\vspace{-0.2in}
\label{fig:1} 
\end{figure}

Our current work builds on previous papers \cite{bacinoglu2015age, yates2015lazy,wu2017optimal_ieee,sun2017update,infocom_arxiv,infocom_w,yinsun, leng2019age, ceran2019average} with the added new direction of research due to intermittent availability of energy for sequential non-preemptive operations in the device. In particular, we consider an energy harvesting node sending status updates to a receiving node through sensing and communication operations as shown in Fig. \ref{fig:1}. Sensing operation includes all preprocessing performed before synthesizing a status update to be sent to the receiver. Transmission involves modulation, forwarding, relaying type operations performed until the receiver receives the status update. 

In our model, the time it takes for sensing and for a packet to be transmitted are both random variables that have general distributions, are independent over time, and are also independent of each other. A single unit data buffer is available for the transmission server so the system can store one packet during power outage. \textit{Neither the transmission operation nor the sensing operation can be preempted.} Sensing and communication operations are activated by a single energy recharge of the node and new energy cannot be harvesting during operation. Once energy arrives, the device allocates it for sensing or transmission and it performs this decision using a \textit{renewal policy}. We prove that average peak age of information (AoI) at the receiver is minimized by a threshold-based stopping rule that accepts only \textit{young} packets to the transmission server. This result is in full agreement with \cite{kam2018age} that introduces packet deadlines to discard the packets in a single server system for timeliness. In contrast, our result differs from other works on AoI in energy harvesting systems such as \cite{bacinoglu2015age, yates2015lazy,wu2017optimal_ieee,sun2017update,2018information,infocom_arxiv,infocom_w,leng2019age,ceran2019average,yinsun} that show usefulness of ``threshold-based waiting" for average AoI\footnote{We also refer to \cite{bedewy2019age} for comparisons between average AoI and average peak AoI with relation to waiting in multi-source status updating. As another related work, \cite{kavitha2018controlling} suggests optimality of ``always drop new packet" or ``always drop old packet". Our conclusion, on the other hand, favors dropping only those old packets whose instantaneous age are above a threshold.}. We then address average AoI optimization over the considered stopping rules through novel hybrid waiting and thresholding schemes. We combine thresholding and packet management and provide comparisons of the AoI performances. Our numerical results show the improvements in average AoI maintained by hybrid schemes. 

\section{System Model}
\label{sec:Model}

We consider an energy harvesting node with sensing operation followed by a transmission queue as shown in Fig. \ref{fig:1}. The sensing represents the initial operations to generate a status update packet. We assume that there is always a packet to generate reminiscent of the \textit{generate-at-will} type policies considered in the literature but it takes some time to complete generation. The generated packet starts aging as soon as the sensing is activated. Once sensing is completed, a status update packet is forwarded to the transmission queue. 

There is a single data buffer to save the latest arriving packet when the system is in power outage. The transmitter chooses to send the latest arriving update or to discard it and generate a new one. Transmissions are performed one at a time and its duration is a general random variable. Once transmission is completed, the receiver (Rx) has the most recent update. Energy is depleted at the end of the sensing and computing operations. The node therefore knows implicitly when the transmission ends. The operations are not allowed to be preempted. The sensing time for packet generation is independent with a general distribution $f_C(c)$, $c \geq 0$ with well-defined mean $\mathbb{E}[C]$ and second moment $\mathbb{E}[C^2]$. Similarly, the time for a packet to be transmitted is independent with a general density function $f_T(t)$, $t \geq 0$ with well-defined mean $\mathbb{E}[T]$ and second moment $\mathbb{E}[T^2]$. 

The energy arrives to the EH node one at a time according to a Poisson process of rate $\lambda$. As soon as energy arrives, it is used to activate \textit{either} sensing \textit{or} transmission. If both servers are idle at the time of an energy arrival, then the energy is allocated to sensing and a new status update packet generation process starts. During these operations, new energy arrivals are ignored as preemption is not possible. Once the operation ends, energy is depleted instantaneously and generated packet is stored in the data buffer. After a new energy arrives, the node has to decide whether to send the existing status update or to generate a new update. 

This model is inspired by generate-at-will type works and addresses intermittent energy availability in various types of energy harvesting batteryless sensors such as \cite{hester2017timely,lucia2017intermittent}. The resulting problem is new to the best of our knowledge as the temporal dependence of the sensing and transmission operations due to intermittent availability of energy has not been addressed in the literature. 

\begin{figure}[!t]
\centering{
\hspace{-0.0cm} 
\includegraphics[totalheight=0.20\textheight]{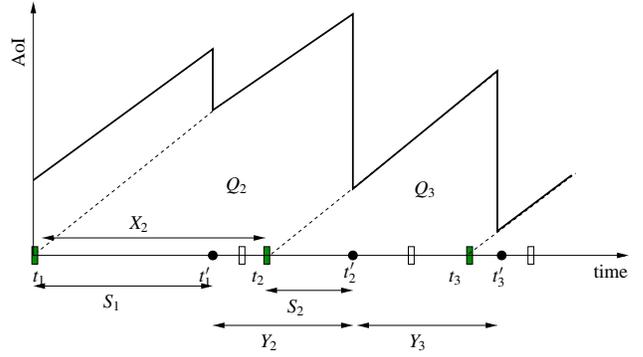}}\vspace{-0.15in}
\caption{\sl Example AoI evolution for general transmission schemes.}
\label{fig:2} 
\vspace{-0.2in}
\end{figure}

We use $t_i$ to denote the packet $i$'s start of generation through sensing, and $t_i'$ to denote the time stamp of the event that the transmission of packet $i$ (if selected for service) is completed. We index the packets that are successfully sent to the receiver and simply ignore those that are discarded. Age of Information (AoI) is measured by the difference of the current time and the time stamp of the latest packet at the receiver: 
\begin{align}
\Delta(t)=t-u(t)
\end{align}
$u(t)$ is the time stamp of the latest received packet at time $t$. We now consider the AoI evolution for a general transmission scheme. We provide a sample AoI evolution curve in Fig. \ref{fig:2}. We assume packet 1 starts aging in the first server at time $0$ when its preprocessing starts. Once it is finished and the subsequent energy arrives, the node takes a favorable decision so that the packet enters the transmission server right away and its service ends at $t_1'$. Then, next energy arrives and it is used to activate sensing (preprocessing) for the next update but after the following energy arrival, the node does not yield a favorable decision and hence the packet is dropped (shown as a hollow rectangle in Fig. \ref{fig:2}). The next energy arrival restarts the preprocessing and this time the node takes a favorable decision and the update is taken to the transmission server. This update's service ends at $t_2'$. Note that dropped packets are not indexed as we count only accepted packets. 

Our focus is on \textbf{renewal policies} where decision to transmit or discard a packet is taken based on observations starting from the last transmission with randomization allowed. From a system operation perspective, the clock (digital counter) at the node is zeroed once a packet is transmitted and it is restarted to count right away. Thus, decisions after $t_i'$ are assumed independent of the past before $t_{i}'$. We assume the same randomized policy is applied after $t_i'$ irrespective of the index $i$.

We define the areas $Q_i$ under the triangular regions of the AoI curve as shown in Fig. \ref{fig:2}. Then, define $Y_i$ as the length of time interval between the departures of packets $i-1$ and $i$ and $S_i$ as the system time for packet $i$. We also define $X_i$ as the length of time interval between the arrivals of packets $i-1$ and $i$ as shown in Fig. \ref{fig:2}. We observe the following
\begin{align}
\mathbb{E}[\Delta]=\lim_{t \rightarrow \infty} \frac{N_t}{t} \frac{1}{N_t} \sum_{i=1}^{N_t} Q_i = \lambda_e \mathbb{E}[Q]
\end{align}
where $\lambda_e=\lim_{t \rightarrow \infty} \frac{N_t}{t}$. It is observed in Fig. \ref{fig:2} that
\begin{align}
Q_i &= \frac{1}{2}\left( (S_{i-1} + Y_i)^2 - S_i^2 \right) \\ &= \frac{1}{2} \left( S_{i-1}^2 + Y_{i}^2 + 2S_{i-1}Y_i - S_i^2\right)
\end{align}
Note that $S_{i-1}$ and $Y_i$ are independent since the decision of a packet being discarded or sent is taken independent of events earlier than the reception of the last packet. Similarly, we note that $S_{i-1}$ and $S_i$ are identically distributed for the \textit{renewal policies}. Hence, we have
\begin{align}
\mathbb{E}[Q_i]=\frac{1}{2}\mathbb{E}[Y_i^2] + \mathbb{E}[S_{i-1}]\mathbb{E}[Y_i]
\end{align}
Since the system is ergodic, we will work with the generic variables for inter-arrival time $X$, inter-departure time $Y$ and system time $S$. We, therefore, have the average AoI as:
\begin{align}\label{aaoi}
\mathbb{E}[\Delta]&=\lambda_e\left(\frac{1}{2}\mathbb{E}[Y^2] + \mathbb{E}[S]\mathbb{E}[Y]\right) = \frac{\mathbb{E}[Y^2]}{2\mathbb{E}[Y]} + \mathbb{E}[S]
\end{align}
where $\lambda_e = \frac{1}{\mathbb{E}[X]} = \frac{1}{\mathbb{E}[Y]}$ as $\mathbb{E}[X]=\mathbb{E}[Y]$. We similarly have the average peak AoI: 
\begin{align}\label{paoi}
\mathbb{E}[\Delta^{p}]=\mathbb{E}[X]+\mathbb{E}[S]=\mathbb{E}[Y]+\mathbb{E}[S]
\end{align}

\section{Average Peak AoI Minimization}

In this section, we consider average peak AoI minimization in the class of renewal policies mentioned in the previous section. Note that due to the renewal property of the considered policies, we express $Y$ and $S$ as
\begin{align}\label{bir}
Y &= I_0 + \sum_{k=1}^{\tilde{n}} C_k + I_k + T \\
S &= C_{\tilde{n}} + I_{\tilde{n}} + T \label{iki}
\end{align} 
where $\tilde{n}$ is a stopping time with respect to $\{C_i+I_i\}$ and $I_0$; that is, decision to set $\tilde{n}=n$ is taken based on the realizations of $\{C_i+I_i\}_{i=1}^n$ and $I_0$. We will indeed see that ignoring $I_0$ does not cause any loss of optimality; still, we assume that $\tilde{n}$ depends on $I_0$ for now. We call $C_i + I_i \triangleq A_i$ and $\tilde{n}$ is conditioned on natural filtrations $\mathcal{F}^{A}_i, i \in \mathbb{Z}_{>0}$ with $\mathcal{F}^{A}_i=\sigma(A_k, k\leq i)$ and $\mathcal{F}^{I}_0=\sigma(I_0)$ of the i.i.d. sequence $\{A_i\}$ and independent random variable $I_0$. By Wald's identity \cite{gallager2012discrete}:
\begin{align} \label{pkl}
\mathbb{E}[\sum_{k=1}^{\tilde{n}} A_k | I_0] = \mathbb{E}[\tilde{n}|I_0]\mathbb{E}[A]
\end{align}
since $A_i$ are independent of $I_0$ and taking expectations on both sides of (\ref{pkl}), we have $\mathbb{E}[\sum_{k=1}^{\tilde{n}} A_k] = \mathbb{E}[\tilde{n}]\mathbb{E}[A]$. We therefore have the average peak AoI
\begin{align}\nonumber
\mathbb{E}[\Delta^p] = \frac{1}{\lambda} +2 \mathbb{E}[T] + \mathbb{E}[\tilde{n}]\mathbb{E}[A] + \mathbb{E}[A_{\tilde{n}}]
\end{align}
We would like to solve the following problem
\begin{align} \label{prob}
\inf_{\tilde{n} \in \mathcal{M}} \mathbb{E}[\Delta^p]
\end{align}
where $\mathcal{M}$ is the space of all stopping times with well-defined mean values over the filtrations $\mathcal{F}^{A}_i$, $i \in \mathbb{Z}_{> 0}$ and $\mathcal{F}^{I}_0$. 

Before we move on, we lay out the following observations about the objective function $\mathbb{E}[\tilde{n}]\mathbb{E}[A] + \mathbb{E}[A_{\tilde{n}}]$ and the stopping rule $\tilde{n}$. In here, the second term $\mathbb{E}[A_{\tilde{n}}]$ could be made smaller by searching for a smaller value of $A_i$ while the first term $\mathbb{E}[\tilde{n}]\mathbb{E}[A]$ increases as a result of this action. There is a tension between optimizing these two terms. We next state our main result in the following theorem:
\begin{Theorem}\label{thm1}
Optimal stopping rule that minimizes average peak AoI $\mathbb{E}[\Delta^p]$ is to stop at the first instance when $A_i < W_{th}$ where $W_{th}$ is the unique solution of the following equation:
\begin{align}\label{elv}
W_{th} = \frac{\mathbb{E}[A]}{\mbox{Pr}(A<W_{th})} +  \mathbb{E}[A | A<W_{th}]
\end{align}
In particular, optimal stopping time is independent of $I_0$ and past realizations of $A_i$. 
\end{Theorem}
\begin{Proof}
If the node decides to stop at time $n$ when $A_n=x$, the cost paid is $I_0 + n \mathbb{E}[A] + x$. If, on the other hand, the node keeps on searching another $A_n$ to stop, then on average with respect to $\{A_k\}_{k>n}$ it has to pay the cost
\begin{align}
I_0 + n \mathbb{E}[A] + c_{x}
\end{align}
to observe a value of $A_{k>n}$ smaller than $x$ where $c_{x}=\mathbb{E}[\tilde{n}_x]\mathbb{E}[A] + \mathbb{E}[A | A<x]$ (using Wald's identity) and $\tilde{n}_x$ is the corresponding stopping time when $A_k$ is below $x$. Then, due to optimal stopping criterion in discrete time by Wald Belmann equation (see e.g. \cite{shiryaev2007optimal}), the optimal stopping time is when instantaneous total cost is smaller than the total expected cost for achieving a smaller $x$ conditioned on current state (in here the state is $(n,x)$). Comparing the two costs, we conclude that when $A_i$ hits $A_i=x$ for $x$ that satisfies the following:
\begin{align} \label{ineq}
x \leq c_{x} \triangleq \frac{\mathbb{E}[A]}{\mbox{Pr}(A<x)} +  \mathbb{E}[A | A<x]
\end{align}
then, the search must be stopped and existing packet must be sent. This argument already shows that $I_0$ and past realizations of $A_i$ are not used in the decision to optimally stop the search. We note that $c_x$ is a continuous function of $x$. Thus, $x \leq c_x$ is a closed set. Since $I$ is exponentially distributed, $A$ has a well defined density $f_A$ with no point masses (even when $f_C$ has point masses). Additionally, $c_x$ is differentiable almost everywhere. For $x=0$, $c_x = \infty$ and  for $x \rightarrow \infty$, $c_{x} \rightarrow 2 \mathbb{E}[A] < \infty$. The first derivative of $c_x$ crosses $0$ only once as the following holds for $x > m_{*}$:
\begin{align*}
\frac{d}{dx} c_x = \frac{x f_A(x)\int_{m_{*}}^x f_A(\alpha)d\alpha - f_A(x)\left(\mathbb{E}[A]+\int_{m_{*}}^x \alpha f_A(\alpha)d\alpha\right) }{\left(\mbox{Pr}(A<x)\right)^2} 
\end{align*}    
where $m_{*}=\inf\{x: f_C(x)>0\}=\inf\{x: f_A(x)>0\}$. We rearrange the nominator of this fraction as
\begin{align*}
f_A(x)\left(\int_{m_{*}}^x (x-\alpha)f_A(\alpha)d\alpha - \mathbb{E}[A]\right)
\end{align*}
Since $f_A(x) > 0$ for all $x > m_{*}$, it suffices to show that the factor $\left(\int_{m_{*}}^x (x-\alpha)f_A(\alpha)d\alpha - \mathbb{E}[A]\right)$ crosses zero only once for $x \geq m_{*}$. This holds since $\int_{m_{*}}^x (x-\alpha)f_A(\alpha)d\alpha$ is monotone increasing and takes value $0$ at $x=m_{*}$. Finally, we note that the following equation is equivalent to (\ref{elv}) (with $W_{th}$ replaced with $x$ and lower limit of integral set to $m_{*}$ as the integral from $0$ to $m_{*}$ is zero): 
\begin{align} \label{ff} \int_{m_{*}}^x (x-\alpha)f_A(\alpha)d\alpha - \mathbb{E}[A] = 0 \end{align} 
Hence, there is a unique solution to (\ref{elv}) that coincides with the unique minimizer of $c_x$ over $x \geq m_{*}$. We conclude that the inequality in (\ref{ineq}) is satisfied with equality for the threshold $W_{th}$ and the optimal stopping set is $\{x: m_{*} \leq x \leq W_{th} \}$.
\end{Proof}

\begin{remark}
The fact that dependence of the stopping rule on the initial power outage time $I_0$ not yielding improved average peak AoI can be extended further. Currently, we assume decisions to send or discard a packet are independent of events before the transmission of the latest update. If we allow dependence of the transmission schemes on earlier times (as in, e.g., seminal papers \cite{sun2017update,yinsun}) while keeping the renewal and ergodicity needed for convergence, this enlarged policy space does not enable strict improvement in average peak AoI. As the expressions in (\ref{paoi}) and following ones (\ref{bir})-(\ref{iki}) remain unchanged in the enlarged space of policies, we can extend the analysis in the proof of Theorem \ref{thm1} to show that dependence on, e.g., $S_{i-1}$ or $T_{i-1}$, to obtain the stopping time for the $i$th update would not change the cost structure and optimal time to stop. In contrast, the average AoI expression in (\ref{aaoi}) would not hold true under dependence of $Y_i$ and $S_{i-1}$.
\end{remark} \vspace{-0.1in}

\subsection{An Example}
\label{secref}
\begin{figure}[!t]
\centering{
\hspace{-0.3cm} 
\includegraphics[totalheight=0.27\textheight]{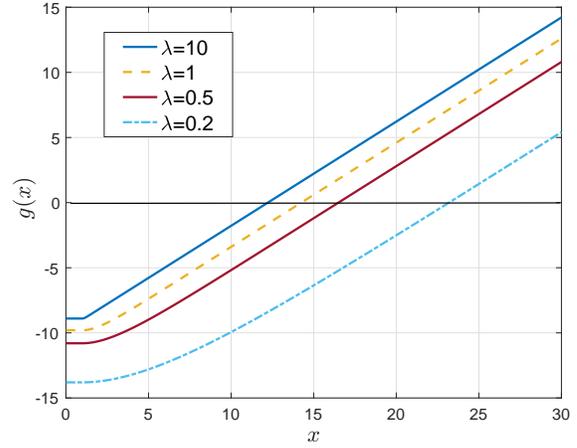}}\vspace{-0.1in}
\caption{\sl Solution of $g(x)=0$ for various $\lambda$. }\vspace{-0.2in}
\label{numres:1} 
\end{figure}

\begin{figure}[!t]
\centering{
\hspace{-0.3cm} 
\includegraphics[totalheight=0.25\textheight]{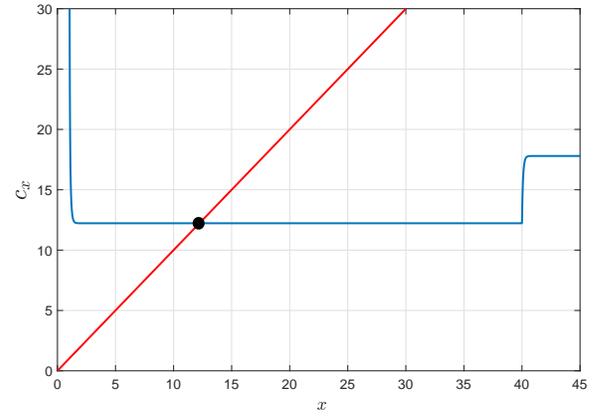}}\vspace{-0.1in}
\caption{\sl The cost $c_x$ for $\lambda=10$ and its minimizer at the fixed point $x=c_x$. }\vspace{-0.2in}
\label{numres:4} 
\end{figure}

\begin{figure}[!t]
\centering{
\hspace{-0.3cm} 
\includegraphics[totalheight=0.27\textheight]{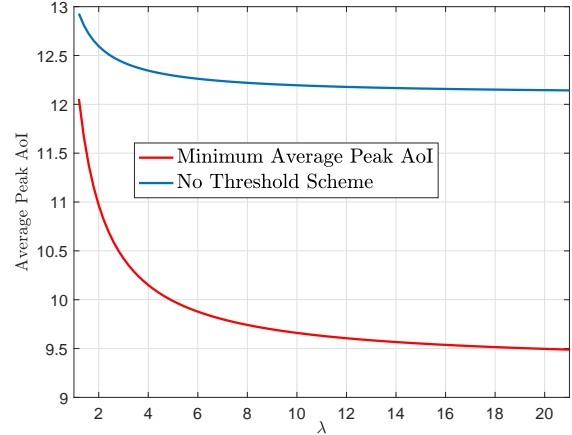}}\vspace{-0.15in}
\caption{\sl Minimum average peak AoI with respect to energy arrival rate $\lambda$ compared to that achieved by benchmark no threshold scheme. }\vspace{-0.2in}
\label{numres:3} 
\end{figure}

We consider a binary valued sensing time $C_i \in \{m_1, m_2\}$ with probabilities $p_1, p_2>0$. In this case, $\mathbb{E}[C]=p_1m_1+p_2m_2$, $\mathbb{E}[A]=\mathbb{E}[C]+\frac{1}{\lambda}$ and $\mathbb{E}[C^2]=p_1m^2_1+p_2m^2_2$. We have $f_A(x)=p_1\lambda e^{-\lambda(x-m_1)}u(x-m_1) + p_2\lambda e^{-\lambda (x-m_2)}u(x-m_2)$ for $x\geq 0$. Here, $u(.)$ is the unit step function. We get: 
\begin{align}\nonumber
\int_0^x f_A(\alpha)d\alpha &= p_1(1-e^{-\lambda (x-m_1)})u(x-m_1) \\ &\hspace{0.2in}+ p_2(1-e^{-\lambda (x-m_2)})u(x-m_2) \\ \nonumber
\int_0^x \alpha f_A(\alpha)d\alpha &= p_1(m_1 + \frac{1}{\lambda} - (x + \frac{1}{\lambda})e^{-\lambda (x-m_1)}) u(x-m_1) \\ &\hspace{-0.5in} + p_2(m_2 + \frac{1}{\lambda} - (x + \frac{1}{\lambda})e^{-\lambda (x-m_2)}) u(x-m_2)
\end{align}
Then, we plot $g(x)=\int_{m^{*}}^x (x-\alpha)f_A(\alpha)d\alpha - \mathbb{E}[A]$, the left hand side of (\ref{ff}), as a function of $x \geq m_{*}$. We take $m_{*}=m_1=1$, $m_2=40$, $p_1=0.8$, $p_2=0.2$ resulting in $\mathbb{E}[C]=8.8$. We observe in Fig. \ref{numres:1} that $g(x)$ crosses 0 at a single point (decreasing for increasing $\lambda$). We also observe in Fig. \ref{numres:4} that the unique minimizer of the cost $c_x$ is the unique fixed point $x=c_x$ (showing for $\lambda=10$). Then, we consider average peak AoI vs. $\lambda$. We fix $\mathbb{E}[C]=5$ so that $\mathbb{E}[A]=5+\frac{1}{\lambda}$. As a benchmark, a no threshold policy is obtained by setting $W_{th}=\infty$. We observe in Fig. \ref{numres:3} that the gains obtained by the optimal policy is more significant as $\lambda$ is increased.

\begin{figure}[!t]
\centering{
\hspace{-0.3cm} 
\includegraphics[totalheight=0.28\textheight]{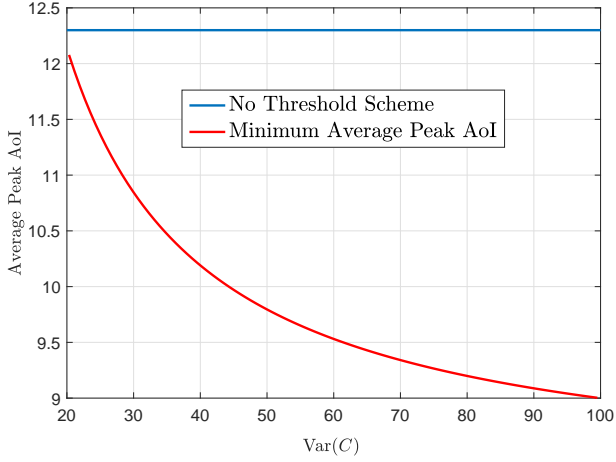}}\vspace{-0.2in}
\caption{\sl Minimum average peak AoI with respect to variance of $C$ compared to that achieved by benchmark no threshold scheme. }\vspace{-0.2in}
\label{numres:2} 
\end{figure}

Next, we fix $\lambda=10$ and $\mathbb{E}[C]=5$ so that $\mathbb{E}[A]=5.1$. We also set $\mathbb{E}[T]=1$. Note that other moments of transmission time $T$ has no influence on the average peak AoI. In this case, the no threshold policy achieves average peak AoI $\mathbb{E}[\Delta^p]=12.3$. Now, we see that variability in $C_i$ enables a more significant improvement in average peak AoI by applying the optimal threshold. We fix $m_1=1$ and parametrize $m_2$ and $p_2$ for fixed mean and increasing variance. Let the parameter be $\theta>0$ so that $m_2=10+\theta$. Then, $(1-p_2)+(10+\theta)p_2=5$ and we have $p_2=\frac{4}{9+\theta}$. We observe in Fig. \ref{numres:2} the minimum average peak AoI compared with that achieved by no thresholding scheme. As the variance of sensing time $C$ is increased under fixed mean, the minimum average peak AoI gets significantly smaller. The search time for a small $A_i$ decreases once the probability mass on $m_2$ shrinks and conditional mean of $A_i$ also decreases as $m_1$ moves closer to $0$. Note also that our observation compares with the theme of the work \cite{talak2018can} for preemptive schemes in the affirmative direction: More determinacy in the sensing time yields larger average peak AoI. 

\section{Stopping Rules for Improved Average AoI}
\label{sec:eval}

In this section, we consider average AoI and how we can improve it by judicious stopping rule design especially using insights obtained from Theorem \ref{thm1} for average peak AoI. Let us start with an arbitrary stopping rule with respect to $\{C_i+I_i\}$ with well-defined mean values over the filtration $\mathcal{F}^{A}_i$, $i \in \mathbb{Z}_{> 0}$. Working on the expression for average AoI in (\ref{aaoi}), we have the following
\begin{align}
\mathbb{E}[Y]&=\frac{1}{\lambda} + \mathbb{E}[T] + \mathbb{E}[\tilde{n}](\mathbb{E}[C] + \frac{1}{\lambda}) \\ \nonumber
\mathbb{E}[Y^2]&= 2\left(\mathbb{E}[T] + \frac{1}{\lambda}\right)\mathbb{E}[\tilde{n}](\mathbb{E}[C] + \frac{1}{\lambda}) + \frac{2\mathbb{E}[T]}{\lambda} \\ &\ + \mathbb{E}[\tilde{n}](\mathbb{E}[C^2]-\mathbb{E}^2[C] + \frac{1}{\lambda^2}) + \frac{2}{\lambda^2} + \mathbb{E}[T^2] \nonumber \\ &\ + \mathbb{E}^2[\tilde{n}](\mathbb{E}^2[C] + \frac{1}{\lambda^2} + \frac{2 \mathbb{E}[C]}{\lambda}) \\
\mathbb{E}[S]&=\mathbb{E}[C_{\tilde{n}}+I_{\tilde{n}}]+\mathbb{E}[T]
\end{align}
and consequently
\begin{align}\nonumber
\mathbb{E}[\Delta] = &\frac{\frac{1}{\lambda^2} + \mbox{Var}(T) + \mathbb{E}[\tilde{n}](\mbox{Var}(C) + \frac{1}{\lambda^2})}{2(\frac{1}{\lambda} + \mathbb{E}[T] + \mathbb{E}[\tilde{n}](\mathbb{E}[C] + \frac{1}{\lambda}))} + \frac{1}{2\lambda} + \frac{3}{2}\mathbb{E}[T] \\&\hspace{-0.1in} + \frac{\mathbb{E}[\tilde{n}]}{2}(\mathbb{E}[C] + \frac{1}{\lambda}) + \mathbb{E}[C_{\tilde{n}}+I_{\tilde{n}}] \label{exp}
\end{align}

\subsection{Discussion}
\label{discussion}

We first observe that the average AoI is dependent on the mean value of the stopping time $\mathbb{E}[\tilde{n}]$ as well as the end state $\mathbb{E}[C_{\tilde{n}}+I_{\tilde{n}}]$. This is suggestive of a similar treatment of the problem to that of average peak AoI minimization in (\ref{prob}). The current form makes it challenging to get exact solution of the average AoI minimization over the considered set of stopping rules. While we do not provide a formal treatment of the average AoI minimization here, we observe that the average AoI expression in (\ref{exp}) is in the following form:
\begin{align} \label{kkl}
H(\mathbb{E}[\tilde{n}]) + \mathbb{E}[C_{\tilde{n}}+I_{\tilde{n}}]
\end{align}
where $H(x)= H_1(x) + \frac{x}{2}(\mathbb{E}[C] + \frac{1}{\lambda}) + \frac{1}{2\lambda} + \frac{3}{2}\mathbb{E}[T]$ is a single variable function over $\mathbb{R}^{+}$ with $H_1(x)=\frac{\frac{1}{\lambda^2} + \mbox{Var}(T) + x(\mbox{Var}(C) + \frac{1}{\lambda^2})}{2(\frac{1}{\lambda} + \mathbb{E}[T] + x(\mathbb{E}[C] + \frac{1}{\lambda}))}$. A formal treatment of the average AoI minimization over the renewal policies of interest would be easier if (\ref{kkl}) is in the form of $\mathbb{E}[H(\tilde{n})] + \mathbb{E}[C_{\tilde{n}}+I_{\tilde{n}}]$. Note that this form may be possible by another state definition for the same problem and we leave it for future. Still, we expect that the solution (if any) is in a \textit{nonstationary} form: Stopping depends on both the current time as well as the value of $C_i+I_i$ (unlike the average peak AoI minimizing rule that depends only on $C_i+I_i$). To support this, note that $H(x)$ is concave, monotone increasing if $K<0$ for $K$ defined as:
\[ K \triangleq (\mbox{Var}(T) + \frac{1}{\lambda^2})(\mathbb{E}[C]+\frac{1}{\lambda}) - (\mbox{Var}(C) + \frac{1}{\lambda^2})(\mathbb{E}[T] + \frac{1}{\lambda}) \]
Indeed, under a linear approximation of $H(x)\approx \beta x + r$ for some $\beta > 0$, we are left with minimizing the following:
\[ \beta \mathbb{E}[\tilde{n}] + \mathbb{E}[C_{\tilde{n}}+I_{\tilde{n}}] \]  
This problem is now in the form of average peak AoI minimization in (\ref{prob}), which has a stopping rule of the form $C_i+I_i < x_{th}$ where $x_{th}$ is the unique solution\footnote{We can show existence and uniqueness of the solution following the same lines to that presented in the proof of Theorem \ref{thm1}.} of the following fixed point equation
\[ x = \frac{\beta}{\mbox{Pr}(C_i+I_i<x)} + \mathbb{E}[C_i+I_i|C_i+I_i<x]\]
We, then, check if the resulting stopping rule has a mean value that is compatible with the assumed linear approximation. An iterative method could help to improve assumed $\beta$ and we leave this for future versions of the current work.   

On the other hand, if $K \geq 0$, then $H_1(x)$ is convex and decreasing and in this case, $H(x)$ may have a non-zero minimizer. Indeed, stopping early (even if a small $C_i+I_i$ is found early) can incur a large average AoI cost. This occurs when variance of $T$ is large relative to its mean. To see this, let us ignore the term $\mathbb{E}[C_{\tilde{n}}+I_{\tilde{n}}]$ in (\ref{kkl}) (e.g., when the variances are large with respect to means and $H(\mathbb{E}[\tilde{n}])$ is dominant). Then, we have a single variable optimization with minimizer:
\begin{align}
x^{*}=\max\{\sqrt{\frac{K}{\mathbb{E}[C]+\frac{1}{\lambda}}}-\frac{1}{\lambda}-\mathbb{E}[T],0\}
\end{align} 
It is, then, clear that for large $\mbox{Var}(T)$, this optimizer is strictly positive and this suggests that an initial ``waiting" irrespective of the observed $C_i+I_i$ is necessary under assumed condition. 

It is remarkable to observe the roles of variances of $C$ and $T$ in the average AoI expression as we are motivated by \cite{talak2018can} to do so. In particular, $H(x)$ in (\ref{exp}) suggests that for fixed means, selecting the variances of $C$ and $T$ as 0 will minimize it. On the other hand, this conclusion may not apply for the average AoI in (\ref{exp}) due to the additional term $\mathbb{E}[C_{\tilde{n}}+I_{\tilde{n}}]$ since this term could be made smaller in the increasing variance regime as we have seen in Section \ref{secref}. 

\subsection{Proposed Stopping Rules}

In light of our discussion on optimizing average AoI, we propose following stopping rules:
\begin{itemize}
\item \textit{Hybrid stopping rules:} These stopping rules are of the following form: First, wait initial $n_w \geq 0$ steps\footnote{If $n_w$ is not an integer, it is always possible to construct a probabilistic stopping time taking values $\lceil n_w \rceil$ and $\lfloor n_w \rfloor$ with mean $n_w$.} irrespective of the value of $C_i+I_i$ and then perform thresholding $C_i+I_i<W_{th}$ to stop. Here, we leave $n_w$ and $W_{th}$ to be numerically optimized. In this case, we have $\mathbb{E}[\tilde{n}]=n_w+\frac{1}{\mbox{Pr}(C_i+I_i<W_{th})}$ and $\mathbb{E}[C_{\tilde{n}}+I_{\tilde{n}}]=\frac{\int_0^{W_{th}} \alpha f_A(\alpha)d\alpha}{\int_0^{W_{th}} f_A(\alpha)d\alpha}$ with $f_A=f_C*f_I$ where $f_A$ is the probability density of $A=C+I$; $I$ is exponentially distributed with rate $\lambda$. 
\item \textit{POD (Power Outage-Based Discarding):} We also consider a practical threshold based scheme that is in the general hybrid stopping rules. For this scheme, decision is taken based on thresholding the time spent in power outage\footnote{This is especially a good fit for cases when sensing time variance is relatively small and main source of uncertainty is in the time spent in power outage.}. Let $W_{pod}$ denote the threshold. If $I>W_{pod}$, then existing packet is discarded and new one is generated. We have $\mbox{Pr}[I<W_{pod}]=1-e^{-\lambda W_{pod}}$ and $\mathbb{E}[\tilde{n}]=n_w+\frac{1}{1-e^{-\lambda W_{pod}}}$. Note also that $\mathbb{E}[I_{\tilde{n}}]=\frac{1}{\lambda}-\frac{W_{pod}e^{-\lambda W_{pod}}}{1-e^{-\lambda W_{pod}}}$ while $\mathbb{E}[C_{\tilde{n}}]=\mathbb{E}[C]$.
\end{itemize}

\subsection{Numerical Results}

In this section, we provide comparisons of average AoI performances of the proposed schemes. We consider a binary valued sensing time $C_i \in \{m_1, m_2\}$ with corresponding probabilities $p_1, p_2>0$. For this case, $\mathbb{E}[C]=p_1m_1+p_2m_2$ and $\mbox{Var}(C)=p_1m_1^2+p_2m_2^2-\mathbb{E}^2[C]$. We set $m_1=1$ and $\mathbb{E}[C]=5$ while we leave $m_2$, $p_2$ as variables to be determined. We also set $\mathbb{E}[T]=1$ while leaving $\mbox{Var}(T)$ as variable.

\begin{figure}[!t]
\centering{
\hspace{-0.3cm} 
\includegraphics[totalheight=0.28\textheight]{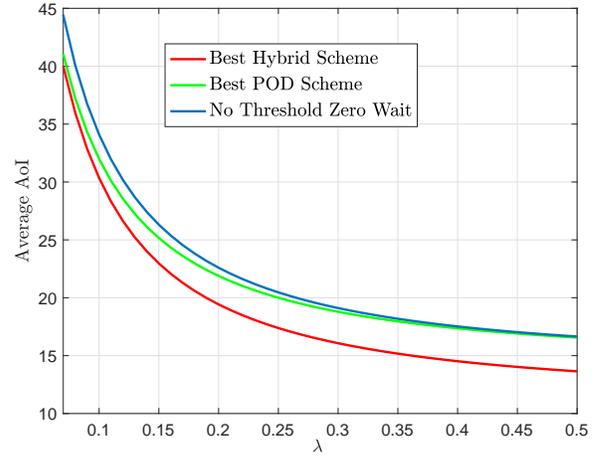}}\vspace{-0.1in}
\caption{\sl Average AoI performances of hybrid stopping, POD and no threshold zero wait schemes with respect to energy arrival rate $\lambda$. }\vspace{-0.2in}
\label{numres:5} 
\end{figure}

\begin{figure}[!t]
\centering{
\hspace{-0.3cm} 
\includegraphics[totalheight=0.28\textheight]{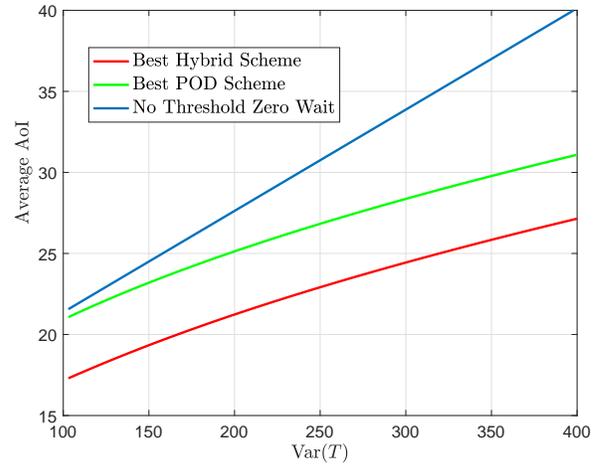}}\vspace{-0.15in}
\caption{\sl Average AoI performances of hybrid stopping, POD and no threshold zero wait schemes with respect to variance of $T$. }\vspace{-0.25in}
\label{numres:7} 
\end{figure}

We first take $m_1=1$, $m_2=21$, with $p_1=0.8$, $p_2=0.2$ resulting in $\mathbb{E}[C]=5$ and $\mbox{Var}(C)=64$. Additionally, we fix variance of $T$ as $\mbox{Var}(T)=1$. In Fig. \ref{numres:5}, we observe the average AoI performances of the proposed hybrid thresholding and no threshold zero wait schemes. This plot represents the general trend: When $\lambda$ is small especially compared to $\frac{1}{\mathbb{E}[C]}$, the POD scheme starts to make significant improvement in the average AoI performance. In this particular case, best POD scheme appears to be almost as good as the best hybrid scheme for smaller $\lambda$ while the difference is significant for large $\lambda$. In this regime of $\lambda$, we observe the other extreme that the best POD scheme has almost identical performance to no threshold zero wait scheme. Note that in this case we have $K<0$ and $H(x)$ is monotone increasing. We observe in our numerical results that optimal selection of waiting $n_w$ is zero in both POD and hybrid scheme. Next, in Fig. \ref{numres:7}, we consider the same setting with variable $\mbox{Var}(T)$ while we fix $\lambda=1$. It is observed that as $\mbox{Var}(T)$ is increased the improvements in average AoI by POD and hybrid schemes become more significant. Introducing non-zero waiting $n_w$ starts to help in the increasing variance regime. These observations support our discussion in Section \ref{discussion}.

\begin{figure}[!t]
\centering{
\hspace{-0.3cm} 
\includegraphics[totalheight=0.29\textheight]{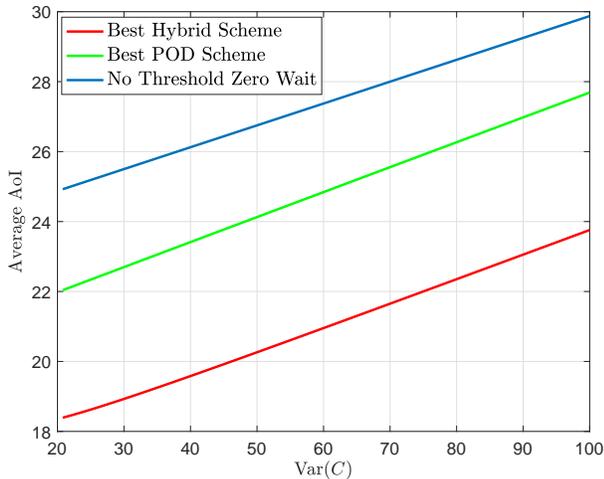}}\vspace{-0.2in}
\caption{\sl Average AoI performances of hybrid stopping, POD and no threshold zero wait schemes with respect to variance of $C$. }\vspace{-0.25in}
\label{numres:6} 
\end{figure}

Finally, we change the variance of sensing time $C$ by changing the position of $m_2$ and its probability mass $p_2$ as $m_2=10+\theta$ and $p_2=\frac{4}{9+\theta}$ for $\theta>0$ so that expected value of $C$ is kept at $5$ while its variance is increased with increasing $\theta$. We set the variance of $T$ as $\mbox{Var}(T)=200$ and arrival rate is kept at $\lambda=1$. In Fig. \ref{numres:6}, we observe the order of AoI performances of best hybrid and POD schemes as well as no threshold zero wait scheme. As predicted by our earlier discussions, average AoI improvements get significant for large variances. One major difference we observe here in average AoI plots compared with average peak AoI is that optimizing the former yields an increasing figure with the variances of $C$ and $T$ but optimizing the latter one yields a decreasing figure. \vspace{-0.05in}

\section{Conclusions}
\label{sec:Conc}

We considered a novel \textit{intermittent status updating} model through sensing and transmission operations. Each operation costs a single energy recharge of the node and one of them is activated at each energy arrival instant. It is not possible to harvest energy during operation. Once Poisson energy is harvested, the node decides whether to activate sensing to generate a new update or transmission to send the existing update (if any) to the receiver. We proved that average peak age of information (AoI) at the receiver is minimized by a threshold-based rule that accepts only \textit{young} packets to the transmission server. We then addressed average AoI optimization over the considered stopping rules through novel hybrid waiting and thresholding schemes. We provided numerical results showing the improvements in average AoI through hybrid schemes.

\end{document}